%% file: main.tex
\documentclass[10pt, conference, letterpaper]{IEEEtran}
\IEEEoverridecommandlockouts
\usepackage[table]{xcolor}
\usepackage{tikz}
\usepackage{amsmath, amssymb}
\usepackage{algorithmicx}
\usepackage{algpseudocode}
\usepackage{listings}
\usepackage{makecell}
\usepackage{textcomp}
\usepackage{xspace}
\usepackage{multirow}
\usepackage[ruled,vlined]{algorithm2e}
\usepackage{footnote}
\usepackage{threeparttable}
\usepackage{color}
\usepackage{booktabs}
\usepackage{subcaption}
\usepackage{ulem}

\definecolor{lightgray}{gray}{0.9}

\usepackage{filecontents}
\usepackage{enumerate}
\usepackage{enumitem}

\usepackage{amssymb}% http://ctan.org/pkg/amssymb
\usepackage{pifont}% http://ctan.org/pkg/pifont
\newcommand{\cmark}{\ding{51}}%
\newcommand{\xmark}{\ding{55}}%

\usepackage{hyperref}
\hypersetup{
    colorlinks=true,
    linkcolor=blue,
    filecolor=magenta,      
    urlcolor=cyan,
}

\newcommand\blfootnote[1]{%
  \begingroup
  \renewcommand\thefootnote{}\footnote{#1}%
  \addtocounter{footnote}{-1}%
  \endgroup
}

\begin{document}
%-------------------------------------------------------------------------------

%don't want date printed
% \date{}

%  make title bold and 14 pt font (Latex default is non-bold, 16 pt)
\title{\Large \bf Ruledger: Ensuring Execution Integrity in Trigger-Action IoT Platforms}
%:\\A Blockchain Approach}

\author{
\IEEEauthorblockN{
Jingwen Fan\IEEEauthorrefmark{1}\IEEEauthorrefmark{4},
Yi He\IEEEauthorrefmark{2}\IEEEauthorrefmark{4},
Bo Tang\IEEEauthorrefmark{1},
Qi Li\IEEEauthorrefmark{2},
and Ravi Sandhu\IEEEauthorrefmark{3}
}
\IEEEauthorblockA{
\IEEEauthorrefmark{1}Information Security Laboratory, Sichuan Changhong Electric Co., Ltd. \\
\IEEEauthorrefmark{2}Institute for Network Sciences and Cyberspace \& Department of Computer Science, Tsinghua University; BNRist \\
\IEEEauthorrefmark{3}Institute for Cyber Security and \linebreak
Department of Computer Science, University of Texas at San Antonio
}
{Email: 
\IEEEauthorrefmark{1}{\{jingwen.fan, bo.tang\}}@changhong.com,
\IEEEauthorrefmark{2}{\{yihe2020, qli01\}}@tsinghua.edu.cn,
\IEEEauthorrefmark{3}ravi.sandhu@utsa.edu
}
}

\def\red#1{#1}
\def\platform{Device Handler Layer}
\def\ourwork{Ruledger\xspace}
\def\blockchain{\textbf{anonymous}\xspace}
% FISCO-BCOS~\cite{fisco-bcos}
%\def\mark#1{\textcolor{red}{#1}}
\def\mark#1{#1}

\thispagestyle{plain}
\pagestyle{plain}

\maketitle

\input{sections/0-abstract.tex}
\input{sections/1-introduction.tex}

\input{sections/2-background}
\input{sections/3-design}
\input{sections/4-security-analysis}
\input{sections/5-implementation}
\input{sections/7-discussion}

\section{Conclusion}
In this paper, we propose \ourwork, a ledger based IoT platform, which is used to ensure the integrity of rule executions in trigger-action based smart home systems. Particularly, \ourwork utilizes applications built upon ledger wallets to honestly record information generated by smart home systems in the ledgers via ledger transactions, and smart contracts automatically verify the authenticity of the information associated with rule executions according to ledger transaction records. We prototype \ourwork with a real trigger-action platform and the experiment results with the prototype demonstrate that \ourwork incurs acceptable overhead for real deployment. 

\section*{Acknowledgement}
The work is supported in part by the National Key R\&D Program of China 
under Grant 2018YFB1800304 and BNRist under Grant BNR2020RC01013. Bo Tang and Qi Li are the corresponding authors of this paper. 

%-------------------------------------------------------------------------------
%\newpage
\bibliographystyle{plain}
\bibliography{main}

\appendices
\input{sections/9-appendix}

%%%%%%%%%%%%%%%%%%%%%%%%%%%%%%%%%%%%%%%%%%%%%%%%%%%%%%%%%%%%%%%%%%%%%%%%%%%%%%%%
\end{document}

%% file: sections/0-abstract.tex
%-------------------------------------------------------------------------------
\begin{abstract}

Smart home IoT systems utilize trigger-action platforms, e.g., IFTTT, to manage devices from various vendors. %These platforms deploy user-defined rules for automation among devices. 
These platforms allow users to define rules for automatically triggering operations on devices. 
However, they may be abused by triggering malicious rule execution with forged IoT devices or events violating the execution integrity and the intentions of the users. To address this issue, we propose a ledger based IoT platform called Ruledger, which ensures the correct execution of rules by verifying the authenticity of the corresponding information. Ruledger utilizes smart contracts to enforce verifying the information associated with rule executions, e.g., the user and configuration information from users, device events, and triggers in the trigger-action platforms. In particular, we develop three algorithms to enable ledger-wallet based applications for Ruledger and guarantee that the records used for verification are stateful and correct. Thus, the execution integrity of rules is ensured even if devices and platforms in the smart home systems are compromised. We prototype Ruledger in a real IoT platform, i.e., IFTTT, and evaluate the performance with various settings. The experimental results demonstrate Ruledger incurs an average of 12.53\% delay, which is acceptable for smart home systems.
\blfootnote{\IEEEauthorrefmark{4} The first two authors contributed equally to this paper.}

\end{abstract}

%% file: sections/1-introduction.tex
%-------------------------------------------------------------------------------
\section{Introduction}
%-------------------------------------------------------------------------------
% 需要添加一些引用
% 
Smart Home Systems\footnote{In this paper, we focus on studying the security of typical IoT systems, i.e., smart home systems. We use IoT systems and smart home systems interchangeably.} and Internet of Things (IoT) devices have proliferated recently. Users use various devices, e.g. the smart lock of Samsung SmartThings~\cite{smartthings-web}, the smart watch of Garmin, as well as bulbs and smoke detectors in their smart home systems. All these facilities can be managed via trigger-action platforms, such as IFTTT \cite{ifttt}, Microsoft Flow \cite{microsoftFlow}, and Zapier \cite{zapier}, and work together to provide different functionalities. %, e.g., ``when there's no one at home, lock all doors and windows and start the security camera to detect intruders". 
Meanwhile, users set up rules in these trigger-action platforms to automate rule executions in their devices. Thus, these platforms are able to manage various devices from different vendors to automatically perform physical or virtual tasks according to the rules. 

Meanwhile, smart home systems suffer severe security threats, %e.g., privacy breach, data breach, and even user safety. %security risks are also introduced endangering user privacy, data security, and even personal safety.
%For example, IFTTT %has serio's are widely used by users to set automation rules which involve users' devices and data, and 
%has various serious issues 
such as privacy leakage~\cite{ifttt_recipts} and violation of rule execution integrity~\cite{ifwhat}. In particular, violating rule execution integrity \mark{becomes} a major concern of smart home security. For example, an attacker can generate a fake location~\cite{peeve} to the trigger-action platform, e.g., IFTTT, so that the platform will trigger the rule to open the smart lock even though nobody is at home. Similarly, if OAuth tokens in the smart home system are leaked or part of the system is compromised, an attacker can easily access user locations and remotely control IoT devices by invoking APIs~\cite{sok, dtap18}.
These security risks in smart home systems have been extensively studied in the literature~\cite{dtap18, situational, peeve, smartauth}. %The root cause of such risks is that the components of the smart home system may be attacked and the attackers can perform overprivileged operations to exploit the rules and devices or violate users' privacy. 
Unfortunately, as shown in Table~\ref{table:cmp}, none of the existing approaches can effectively \mark{protect the} integrity of rule executions in these systems. %the smart home system suffers four types of attacks, the trigger-action APIs attacks which mean the trigger-action platform is attacked and the rule triggering APIs can be exploited, the platform APIs attacks which means the IoT platform is attacked and the user's rules and devices configuration account can be exploited, device APIs attacks which means the devices' remote APIs are exploited,  privacy violations which mean user's privacy from the IoT devices leaks, and event spoofing attacks which means the attackers can exploit the rules via fake trigger events. To defend against these attacks, secure architecture for IoT platform is needed to provide end-to-end protection for rule execution. However, existing works cannot properly prevent the rule exploited. 
For example, ESO~\cite{situational} enforces \mark{situational} access control for trigger-action platforms by encapsulating IoT remote APIs, which cannot avoid fake triggers in trigger-action platforms. %to use the overprivileged APIs directly. 
DTAP~\cite{dtap18} only enables secure token authorization for  trigger-action platforms but cannot \mark{prevent} malicious rule \mark{executions triggered} by IoT platforms themselves. Similarly, Peeves~\cite{peeve} can only prevent event spoofing attacks by verifying physical events in IoT devices. Thus, the current trigger-action based smart home systems still lack protection for \mark{the} integrity of rule execution.   

\begin{table*}[ht]
\caption{Comparison with existing works.}
%\vspace{-0.8in}
\centering
\rowcolors{1}{}{lightgray}
\begin{tabular}{cccccc}
\hline
 & Trigger-Action API Attacks & Platform API Attacks & Device API Attacks & Privacy Violations & Event Spoof \\ \hline
Peeves~\cite{peeve} & \xmark  & \xmark  & \xmark  & \xmark  & \cmark \\ \hline
DTAP~\cite{dtap18}   & \cmark & \xmark  & \cmark & \xmark  & \xmark \\ \hline
ESO~\cite{situational}    &  \xmark & \xmark  & \xmark  & \cmark & \xmark \\ \hline
\ourwork   & \cmark & \cmark & \cmark & \cmark & \cmark  \\ \hline
\end{tabular}
\label{table:cmp}
\end{table*}

In this paper, we systematically study vulnerabilities that enable malicious rule executions in smart home systems.
The root cause of the vulnerabilities is that there does not exist any mechanism to ensure the authenticity and integrity of the messages \mark{transferred} among different components in smart home systems. To this end,  %and discuss possible attacks in different components of the smart home system. These attack surfaces indicate that the key point to ensure the rule execution integrity and prevent the rules exploited is that both the event sources and the rule trigger requests should be validated. Inspiring by the stateless ledger proposed by Gabriel et al.~\cite{ledger}, 
we propose \ourwork, a ledger-based IoT platform that can be integrated with existing platforms, to ensure \mark{the} integrity of rule executions by enabling verification of the information in the smart home systems. In particular, \ourwork utilizes three ledger wallets to guarantee stateful records according to real information, and leverages smart contracts to enable protections for rule configuration and verification of the trigger events and \mark{action} requests according to the stateful ledger records. \ourwork records all \mark{the action} requests and events in the ledger as stateful transaction \mark{logs} so that it can verify the authenticity of all information associated with rule executions based on the \mark{logs} and the \mark{execution} states in the ledger. Therefore, \ourwork can effectively prevent faked event sources of rule triggers and malicious action requests generated from the trigger-action platforms. The security analysis proves that \ourwork ensures the rule execution integrity under various attacks. We prototype \ourwork with the real IoT devices and the mainstream trigger-action platform, i.e., IFTTT, and evaluate the performance of \ourwork. The experimental results demonstrate that the overhead incurred by \ourwork is  acceptable for real deployment. The \mark{increased} delay is about 12.53\%, and the throughput overhead \mark{decreases} by about 6.5\%, \mark{comparing to the baseline systems}.  

{Our contributions can be summarized as follows.}
\begin{itemize}[leftmargin=*]
    \item {We systematically study the vulnerabilities of violating the integrity of rule executions in trigger-action based smart home systems. }% and attacks in smart home systems and introduce \ourwork to address these issues.}
    \item We present \ourwork, a ledger based IoT platform, which verifies the authenticity of the information (events and the corresponding \mark{execution states}) and ensures the integrity of rule executions. %defend against various attacks. 
    \item We develop \mark{state generation and verification algorithms built upon ledgers} and wallets to guarantee that the states of the transactions in a smart home system are properly verified and submitted.
    %and show how we ensure the rule execution integrity via smart contacts and analyze the security based on the possible attacks. }
    \item {We prototype \ourwork and the experiments with real-world IoT platforms  demonstrate its performance.}
    %the usability and the efficiency  with an overall overhead of 14\% and show the performance in concurrent execution with a decrease of x\% in throughput.}
\end{itemize}

% Finally, we implement and deploy our systems in both local devices and cloud hosts to evaluate the performance in different situations. We implement some basic IoT rules on our system and use the trigger-action platforms such as IFTTT to trigger these rules and show the feasibility of our system. It proves that our system can easily integrate with current third-party platforms and can scale to large systems with negligible overhead. 

%% file: sections/2-background.tex
% hy：改到了40行
%-------------------------------------------------------------------------------
\section{Background} %and Problem Statement} %Definition}
%-------------------------------------------------------------------------------
%In this section, we review the key components in smart home systems and review distributed ledgers.

\subsection{Smart Home Systems}
\begin{figure}[t]
    \centering
    \includegraphics[width=0.43\textwidth]{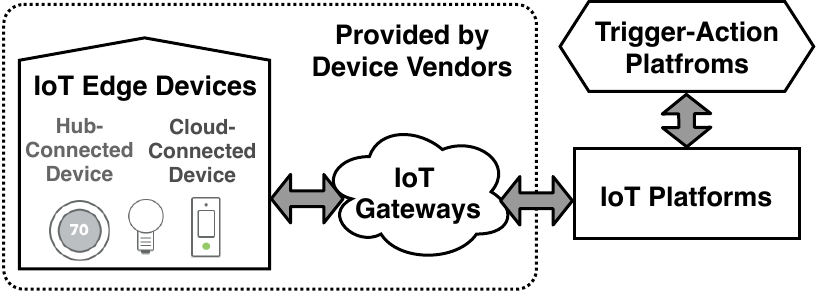}
    % \vspace{-0.06in}
    \caption{The main components of smart home systems. The IoT gateways from device vendors serve as device brokers for IoT platforms. The trigger-action platforms may be third-party.}
    \label{fig:stack}
\end{figure}

Typically, smart home systems leverage trigger-action platforms to achieve inter-device automation by means of predefined rules. Such a trigger-action platform is operated either by a third party, e.g., IFTTT, or as part of a comprehensive IoT platform, e.g., Apple Homekit~\cite{homekit}. A user may configure the automation rules on the trigger-action platform specifying triggering conditions and actions of the corresponding devices. 
% Meanwhile, users can set automation rules in Trigger-Action platforms, and the rules will be automatically distributed to the corresponding devices via the IoT systems and trigger action execution in the IoT devices. 
% with the help of the Smart Home system. From a bottom-up perspective, the typical deploy stack of the Smart Home is shown in Figure~\ref{fig:stack} and contain the following components. 
The data flow in a typical smart home system is illustrated in \mark{Fig.}~\ref{fig:stack} \mark{which contains} three components i.e., IoT edge devices, the IoT platform, and the trigger-action platform. The devices in a home may be from various vendors and \mark{connected to} IoT platforms through the designated IoT gateways \mark{which are usually from different vendors}. \mark{The IoT gateways interact} with each other \mark{through} trigger-action platforms as specified by user needs via OAuth APIs~\cite{oauth}.

\noindent \textbf{IoT Edge Devices and IoT Gateways.} The IoT edge devices, including the corresponding IoT hubs providing wireless connection services, are the physical devices deployed in users' houses, and IoT \mark{gateways are} deployed on the Clouds which provide remote APIs for device control. Some cloud-connected IoT devices with Internet access can directly connect to their online \mark{gateways}, while some hub-connected devices that only have local wireless access (such as Zigbee, Z-Wave, and Bluetooth) need to connect to their IoT \mark{hubs}. %i.e., SmartThing Hubs to reach their gateway.
To remotely control these devices, IoT gateways provided by device vendors usually export APIs that are implemented via HTTP, WebSocket, or MQTT~\cite{mqtt} for device access, and then the smart home systems can automate rule executions in the devices via these APIs.
% via the Hubs. %-connection. Finally, all the devices can be managed by their backend cloud services. 
% which include the gateway service and IoT automation services providing by the IoT Framework.

\noindent \textbf{IoT Platforms.} IoT Platforms provide a programmable framework that wraps all the functionalities and APIs enabled in an IoT device. For example, SmartThings can set Device Handlers and SmartApps for IoT devices and leverage the Groovy scripts to define properties, e.g., event sending and event subscribing interfaces, for a specific device. Usually, IoT platforms, such as SmartThings and HomeKit, can manage both their own IoT devices and third-party devices. Particularly, the third-party devices first need to connect \mark{to} their own gateways provided by the vendors and then communicate with the IoT platforms through the OAuth protocol~\cite{oauth}.

\noindent \textbf{Trigger-Action Platforms.} \red{Smart home systems allow users to set trigger-action rules to automate rule \mark{executions} in devices. Some IoT platforms allow \mark{direct configuration of} %may provide the 
trigger-action rules (e.g. HomeKit), and %setting function and there are also 
many \mark{widely-used} third-party trigger-action platforms, such as IFTTT~\cite{ifttt}, Zapier~\cite{zapier}, and Microsoft Flow~\cite{microsoftFlow} can configure the rules as well. In this paper, for simplicity, we take the most popular trigger-action platform,  i.e., IFTTT, as an example, which receives an HTTP request as a trigger event and then generates HTTP requests to trigger one or more actions.}

% \red{\textbf{TODO: how rules works and how event generated}}
\red{Unlike IFTTT web apps~\cite{ifttt_study}, the IoT trigger-action rules are triggered and executed in IoT devices, and \mark{they are defined} in forms of \mark{conditional expressions, i.e.,} if the trigger event happens then the actions will be executed in IoT devices. A trigger event can \mark{be a} specific event or a status from the IoT device's operation, e.g. the door is opened or a user \mark{arrives} home, and actions are a serial of operations on IoT devices. To set a trigger-action rule, users first need to choose an IoT operation to generate the trigger event and then set some operations as actions.}
% SmartThings leverage Smart Apps to set rules that are similar to trigger-action operations to automate device execution.

\subsection{Distributed Ledgers} %and Blockchain}

%\noindent \textbf{Distributed ledgers}. 
Distributed ledgers, such as blockchain systems, have been widely used to record various transactions across independent systems. 
A transaction is only ever stored when consensus has been reached by the involved parties. The \mark{light-weighted} consensus algorithms, such as practical Byzantine Fault Tolerance (pBFT)~\cite{pbft} and raft~\cite{raft}, are \mark{extensively} used \mark{to achieve data synchronization} in distributed \mark{ledgers}. In this paper, we leverage pBFT to \mark{achieve consensus among} smart home systems, which is efficient in asynchronous environments. Distributed ledgers utilize client-side agents (called wallets) to publish data in a decentralized database. The published data can also be user-defined executable codes called smart contracts~\cite{ethorium} that can specify ledger functionalities with automatic execution on the records. \mark{Note that, a smart contract can be applied to various distributed ledgers that use different consensus algorithms}.

\section{Uncovering Vulnerabilities in Rule Execution}
\label{sec:attacks}
% In this paper, we focus on the home-based IoT situations and discuss the security problems in trigger-action platforms. 
In this section, we systematically study the attacks that trigger malicious rule executions \mark{and violate} the integrity of rule execution in smart home systems. % and analyze how these attacks impact the security of IoT systems. 

\subsection{Threat Model}
% As is shown in Figrue~\ref{fig:stack}, the communication and devices of each component may suffered privilege escalation attacks. When the devices or the communication channels from devices to cloud services are compromised, the attacker can send fake event triggers.

\red{In this paper, we consider two types of typical attacks, i.e., API level attacks and platform/device compromise attacks, that trigger malicious rule execution on benign devices \mark{and} incur the overprivilege problem in these devices. API level attacks access and manipulate the remote APIs of the systems, e.g., devices APIs, trigger-action API, or configuration APIs, by exploiting leaked OAuth tokens or vulnerable APIs, while platform or device compromise attacks directly compromise and control platforms or devices. Note that, we only consider a small part of devices and platforms \mark{being} compromised, which can ensure correct rule execution in benign devices.} 

\red{Particularly, %we take different thread models for the three components. For 
%in platforms, 
we consider that trigger-action platforms are untrusted, which means attackers can construct API level attacks to %and can be compromised by attackers, which means that the attackers can 
arbitrarily create and trigger rules that will be triggered in benign devices. Moreover, attackers can leverage \mark{API} level attacks~\cite{api_level} or compromise platforms to set malicious rules for benign devices in IoT platforms. %, we assume that limit service nodes are compromised as the system is down if the whole service is compromised. Most of the time, we only discuss the API-level attackers. 
Similarly, attackers can compromise edge devices or launch API attacks to trigger rule executions on benign devices.}
%And for the devices and IoT Gateway, we assume that only limited devices are compromised or affected by API-level attacks. We cannot prevent the exploiting of these compromised devices instead we need to defend against these malicious devices to exploit other benign devices via rule execution. }
% Adam Bates: An API-level attacker is able to access or manipulate the state of the smart home through creation and transition of well-formed API control messages.
\red{Note that, %a platform compromised attacker can perform many attacks such as paralyze the whole system. In this paper, we focus on the attacks that violate the rules executing and do not discuss the attacks such as denial of service.
we focus on attacks that violate the integrity of rule  execution, and thus do not consider other \mark{types} of attacks such as denial of service (DoS).}
% IoT platforms and we do not ensure the security of third-party trigger-action platforms and the IoT gateway provided by the vendors. 

%Our designing goal is to ensure the devices controlling security with the least security assumptions. As every component of the IoT stacks may be attacked, we assume a very strong attack model that both API-level attacks (token compromise, vulnerable APIs) and system compromise attacks (OS vulnerabilities exploited)  even if the attackers totally get control of the trigger-action platforms and IoT frameworks. For the IoT Gateway and IoT devices, we assume that attackers can only perform some API-level attacks which can affect the IoT frameworks and limited devices. We do not assume that the IoT Gateway and all the physical devices are completely compromised. And we assume that only limited devices are affected by the attackers so that we can still prevent attackers to perform event spoofing attacks to misuse other benign devices.

\subsection{Violating Integrity of Rule Execution}
\begin{figure}[t]
    \centering
    \includegraphics[width=0.47\textwidth]{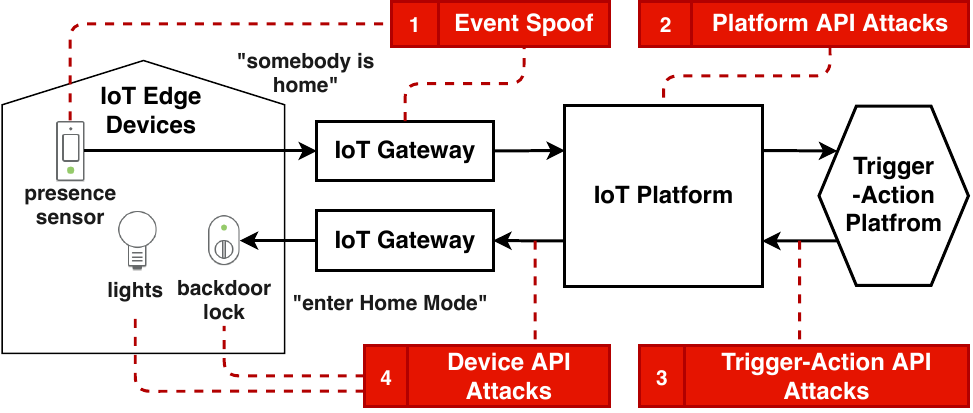}
    % \vspace{-0.1in}
    \caption{Typical attacks against the smart home systems.}
    \label{fig:attack}
\end{figure}

%we study the vulnerabilities of rule  execution in smart home systems. %We find that different components of smart home systems suffer from various attacks. Particular, 
Unlike traditional client-server applications, IoT systems include diversified IoT services and heterogeneous devices. \mark{This makes} the systems extremely complicated and vulnerable to the violation of rule execution integrity \mark{and} abnormal rule executions, which \mark{therefore} incurs the overprivilege problem in benign IoT devices. %Thus, action execution in these systems always suffer to various attacks. 
%%First, an attacker can easily invoke actions in IoT devices by exploiting IoT devices or inside APIs.  
%%%Second, IoT frameworks are easily exploitable to invoke rules in devices or trigger rule execution in trigger-action platforms. %generating fake events. 
%%%Third, trigger-action platforms can directly exploited to trigger rule execution. 
%\subsection{Attacks to the components of IoT Systems}
%fake events are injected into devices or the API services of the devices, which will trigger rule exectution and incur overprivilege 
%and trigger-action platforms 
%and have more attack surfaces as a result, each component of the system may suffer attacks.  Finally, the attacks from the IoT frameworks and trigger-action platforms can lead to the device's arbitrary exploiting.  Also from the device's perspective, if some of the devices or device's API-service are compromised, fake events can be sent to the trigger-action platform to trigger automation rules and cause overprivileged problems. We need to consider three types of attacks or flaws in the IoT system. 
For simplicity, we use the example shown in \mark{Fig.}~\ref{fig:attack} to illustrate how each component of IoT systems may be attacked. The corresponding example rule is \textit{``when the user arrives home, enter home mode and turn on the lamps and open the yard's backdoor"}. 

\red{%There are several attacks that can affect the interactions among IoT platforms, IoT devices (and IoT gateway), and the trigger-action platforms and lead to rule exploit or device exploit. Now we present the detail attacks in different components. 
Now we identify various attacks in different components of smart home systems that can lead to violations of rule execution, i.e., malicious rule execution on benign devices.}
% Now we present the vulnerabilities and attacks in each components.

% by components
% \noindent \textbf{Event Spoofing Attacks.} 
% \red{The rule may be triggered by an fake event in IoT Edge Devices and IoT Gateway.}
\noindent \textbf{Attacks against IoT Devices and IoT Gateways.} 
\red{IoT devices and IoT \mark{gateways} may be compromised by attackers who can \mark{launch} fake events attacks. As \mark{shown} in \mark{Fig.}~\ref{fig:attack}, a rule is triggered by a fake event generated from a compromised device or a compromised IoT gateway that brokers messages to and from the devices\footnote{For simplicity, in this paper, we do not distinguish attacks against IoT \mark{devices} and IoT gateways since the results have the same effect, i.e., spoofing events.}.} It is difficult to prevent attackers from compromising devices and arbitrary abuse of compromised devices. An attacker can easily \mark{conduct} event spoofing attacks in devices (or gateways) to generate fake events, which trigger malicious rule \mark{executions} in \mark{various} benign IoT devices. Let us follow the example shown in \mark{Fig.}~\ref{fig:attack}. An event spoofing attack can easily spoof a state of the presence sensor which tricks the smart home IoT systems and the corresponding collaborative devices, such as lights, backdoor locks, etc., to activate the Home Mode when everyone is away. 

In order to prevent the malicious devices \mark{from} triggering automation rules for other devices, Peeves~\cite{peeve} utilizes an \mark{off-line} event verification mechanism to ensure the authenticity of events\mark{, which uses} the data of other sensors to check if their status have really changed after the event. However, this approach \mark{cannot} be applied in trigger-action IoT \mark{systems that} require instant rule execution after an event occurs.
%the changes of other sensors' status.
%really changes while the rules are usually triggered immediately. 
%We use event logs to verify if the event is triggered by legal users and ensure that only legal events can trigger the rules.

% Also when other components are compromised, attackers can arbitrary trigger existing rules without compromise the Trigger-Action platforms and exploit other devices.  Event verification mechanism is needed to ensure that the trigger-action rules can not be triggered by the fake event from attackers when the  API-service is compromised. 

% PEVEE~\cite{peeve} utilizes the status of other sensors and machine learning algorithms to verify if the event sending by devices is authentic.
% Our works try to solve all the problems by ensuring every event is real and the rule can only be triggered by the platform with the correct permission.

\noindent \textbf{Attacks against Trigger-Action Platforms.} 
Third-party trigger-action platforms allow users to use trigger-action rules to automate IoT devices and usually, a token can be used by several APIs and involves multiple rules and devices. The token is often overprivileged and can be used for controlling all devices from a user. When the OAuth token is misused due to the \mark{platform or token being compromised}, the attackers can abuse the rules to perform risky operations~\cite{cross-cloud}. As is shown in \mark{Fig.}~\ref{fig:attack}, the untrusted \mark{trigger-action platforms} may directly exploit the home mode actions without the trigger event of the \mark{user arriving home}.

DTAP~\cite{dtap18} aims to throttle the attacks by enabling a fine-grained short-lived token for each operation specified in a rule to ensure secure rule usage. In this setting, IoT platforms can issue new tokens to block malicious trigger requests. %Unfortunately, 
% bastys2018if

% \textbf{Configuration API Attacks.}
\noindent \textbf{Attacks against IoT Platforms.} 
As IoT platforms are used to configure and execute rules, they need to interact with IoT devices and the trigger-action platforms. The IoT platforms should ensure the security of configuration APIs, which are used to set rules and manage user accounts and gateway APIs, as well as detecting event spoof attacks generated from the devices and \mark{preventing} malicious requests from the trigger-action platforms.  %which are used to manipulate devices. 
If \mark{attackers exploit} the configuration APIs or the devices APIs, they can either trigger the malicious behaviors by modifying the trigger condition or trigger rule executions in the devices directly. Typically, rule execution is performed in the IoT platform, which invokes the Gateway APIs to control devices via the OAuth protocol, and all tokens of various devices are stored in the same place in the IoT platform. When the IoT platform is under the API level attacks (i.e., by using leaked tokens or vulnerable APIs) or is compromised, the attacker can easily abuse the \mark{device APIs}. %The token security problem can be solved by techniques such as hardware-backed secure storage or TEE. And we also need to prevent the IoT platforms from arbitrary exploiting the gateway APIs.

In order to prevent API abuse, DTAP~\cite{dtap18} develops an OAuth token protection mechanism for the trigger-action platform. However, it does not protect the IoT platform itself and cannot prevent rule \mark{tampering attacks that leverage} vulnerable \mark{configuration} APIs, e.g., manipulating user account settings, rule configurations, or the scripts that implement rules. %when the platform configuration APIs are compromised.
Thus, DTAP is unable to prevent platform API exploitation that triggers malicious rule execution. 

%% file: sections/3-design.tex
\section{Overview of \ourwork}   \label{sec: design}
\def\scene{\textit{"heart rate alert"}}
In this section, we present an overview of \ourwork, which ensures execution integrity in smart home systems by establishing ledger based event and rule verification, and discuss the challenges in the design. 
%addresses the challenges by developing a ledger based IoT framework. 
%and illustrate how we address the challenges of executing integrity by utilizing a ledger-based IoT framework that can guarantee the security requirements of G.1 to G.4. 

\subsection{Overview} \label{subsec: overview}
\begin{figure}[t]
    \centering
    \includegraphics[width=0.48\textwidth]{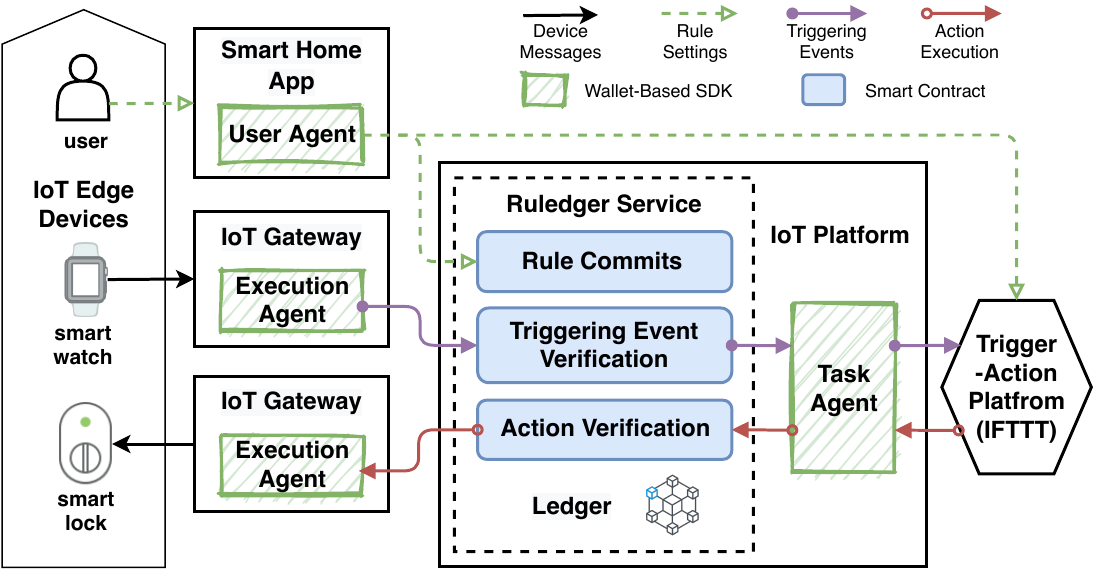}
    % \vspace{-0.1in}
    \caption{The architecture of \ourwork}
    \label{fig:overview}
\end{figure}

In this paper, we propose \ourwork, a distributed ledger-based IoT platform, which aims to ensure the integrity of rule execution by verifying the authenticity and consistency of the generated information in smart home systems. In particular, \ourwork utilizes wallet agents to enforce  configurations and interactions among IoT Apps, IoT gateways, IoT platforms, and trigger-action platforms via transactions on the ledger. \mark{Different platforms can use APIs provided in the SDK of the wallet agent to realize interactions among different platforms. Ruledger} can be integrated with existing third-party trigger-action platforms, which prevents various attacks of violating the rule execution integrity in smart home systems with trigger-action features. Note that, in \ourwork, we do not consider the API level attacks compromising the private keys in the wallets, which have been addressed by existing technologies~\cite{key_protect1, key_protect2}. \mark{Fig.}~\ref{fig:overview} 
shows the architecture of \ourwork. It consists of the rule commits module, the triggering event verification module, and the action verification module. 

%\vspace{-0.5in}

(i) The \textbf{rule commits} module \mark{guarantees} all configurations are authentic. It consists of a rule commits smart contract in the IoT platform and wallet agents called user agents on the user side. This smart contract is used to set up device information and rule configurations in the ledger according to the information collected from the user agents that are integrated with users' IoT management apps. To configure a specific rule, users need to bind the device and set up the OAuth credential for connecting the corresponding gateway, and then set up trigger-action rules by choosing trigger operations and action operations in the app. All trigger-action rules will be committed to the ledger and exported to the third-party trigger-action platforms by the user agent. 

(ii) The \textbf{triggering event verification} module verifies if the triggering event actually happens. It includes \mark{an} event verification smart contract and wallet agents called execution agents in the IoT gateway. The smart contract verifies if triggering events \mark{are authentic} by checking if the \mark{result of} the trigger operation meets the trigger conditions according to the information reported from the execution agents. When a rule is triggered, each operation in the rule is associated with a standalone program, e.g., each operation in SmartThings devices is implemented \mark{as} a device handler, in a groovy sandbox. The \mark{rule} execution is performed by our execution agent if it meets the trigger condition. All triggering results are submitted to the ledger via transactions, and the execution logs \mark{that bind to} a one-time key are recorded. The event records will be committed \mark{to} the ledger if and only if the trigger transactions and the event logs are consistent, which prevents \mark{event spoofing} attacks using fake execution logs.   

%\red{The event verification module is used to verify if the triggering event really happens. After setting the rules, the trigger operations of the rules keep executing and check if meet the trigger conditions. Base on the sight that a rule consists of several independent operations and each operation can be executed as a standalone program, such as the SmartThings devices' operations are implemented as device handlers, and each of them runs as a script in a groovy sandbox.  In our system, each operation is executed by the independent execution agent then if meet the trigger condition, the triggering result is submitted to the ledger and the execution log is stored with a randomness key to prevent the attackers counterfeit log.   Then the event verification module will check the transaction and the event log to ensure the authenticity of the event. Finally, the event record will be committed to the ledger.}

(iii) The \textbf{action verification} module includes \mark{an} action verification smart contract and task agents. The action verification smart contract verifies if the action should \mark{be executed} by checking \mark{whether} the action requests from the trigger-action platform \mark{are} valid according to the ledger record and the transaction detail from the task agent. After the ledger \mark{receives} an action request from the trigger-action platform, the \mark{action} verification module checks if the condition of the rule is met based on the triggering event records on the ledger. Note that, when a new event record (discussed in ii)  is committed to the ledger, the task agent reads the trigger record from the ledger and sends a trigger request to the trigger-action platform. The corresponding rule is triggered and the trigger-action platform sends action requests to trigger the action operations. These requests are converted to transactions by the task agent and the action verification module \mark{verifies} if the previous corresponding trigger records are authentic. If the previous triggering event records are real and the trigger condition in the rule is met, the action operations will be executed. Meanwhile, these action operation transactions will be committed to the ledger and the execution agent updates with the records of these operations to verify subsequent rules. % to motivate the next rules. 

\subsection{Challenges}
% \red{It is non-trivial to design and implement these modules. There are still some challenges.}
It is challenging to develop \ourwork since the interactions among IoT platforms, trigger-action platforms, and edge devices may suffer various attacks exploiting rules, events or devices. We cannot simply rely on ledger transactions to achieve rule execution integrity. %From a high level, there are still some challenges mainly stemming from (i) how to provide replay resistance and fine-granted tokens for the trigger-action platform (ii) ensure the rule security of the IoT framework (iii) event verification. From the module's perspective, the challenges we face for each part are as follows.
%However, third-party trigger-action platforms such as ITFFF are untrusted~\cite{ifttt_recipts} and the event sources lack verification mechanisms. It means that the interactions among IoT framework, trigger-action platforms, and the devices may suffer attacks and lead to rule exploit or devices exploit. We cannot simply rely on the ledgers to implement the execution integrity. From a high level, there are still some challenges mainly stemming from (i) how to provide replay resistance and fine-granted tokens for the trigger-action platform (ii) ensure the rule security of the IoT framework (iii) event verification. From the module's perspective, the challenges we face for each part are as follows.

\begin{itemize}[leftmargin=*]

    \item \textbf{Authenticity of Rule Configurations.} Trigger-action platforms and IoT platforms usually contain rules generated by various users and a user may also have multiple devices. It is difficult to prevent attackers or malicious users \mark{from generating} malicious configurations. \ourwork should ensure rule executions on correct devices and prevent \mark{overprivileged} operations, especially when the platform is compromised. 
    %The challenge here is that we should ensure that rules should only access the needed devices and the authorization mechanism should prevent any overprivileged operations especially when the platform is compromised.\red{The challenge is to prevent the attackers to perform malicious settings.}
    
    \item \textbf{Preventing Fake Triggering Events.} {The IoT platform should be able to verify if an event actually happens in the device and the rule execution condition is met after an event is triggered. However, it is difficult to achieve this because attackers can exploit devices with fake events by using the devices’ cloud APIs. Moreover, the gateway's OAuth tokens should be protected and the event should be verified when a trigger operation occurs.}
    %IoT platforms should avoid the rule exploiting by the fake triggering events.}
    %\red{For a triggering event, the IoT platform should verify if the event happened in the devices and the rule's execution condition is right as attackers can exploit the devices directly by using the devices' cloud API or send fake events. The gateway service's OAuth token should be protected and the event should be verified when a trigger operation happened. IoT platforms should avoid the rule exploiting by the fake triggering events.}

    \item \textbf{Preventing Malicious Transactions.} \ourwork should prevent malicious transactions generated by unauthorized rule administration in trigger-action platforms, which is difficult to achieve. Since the same OAuth tokens are always shared by multiple rules and devices, malicious transactions can be generated by sending action requests to the task agent with rewind attacks using reused or leaked tokens.

\end{itemize}

\section{Design Details}
In this section, we describe the design details of \ourwork and address the design challenges. 
\subsection{Rule Commits}
% \ourwork adopt role-based access control for rule settings via the smart contact and store the devices and rule configurations in the ledger. 

{We should prevent vertical privilege escalation~\cite{mace} during committing rule configurations. Otherwise, attackers can abuse multiple users' rules and devices when the platform is compromised. To address this issue, we utilize a wallet based user agent, which ensures that users correctly configure devices and rules and manage their accounts. In particular, \ourwork adopts the role-based access control model for users to configure rules via the smart \mark{contract}, which allows users with different permissions to correctly record all configurations in the ledger via the user agents. A sample smart contract for verification can be found in Appendix~\ref{adx:smart_contact}. Similarly, all changes of configurations will be committed via ledger transactions, which are triggered by the user agent as well. %with specified access control polices. %is ensured by the permission of the wallet. 
To achieve this, different user agents with corresponding roles use their private keys indicating different permission levels, e.g., administrator's or normal user's private key. Finally, each transaction operation is verified by the smart contract, which checks the access control list based on the transaction signature signed by the corresponding private key of the user agent. Rather than traditional \mark{ways of} setting rules in the trigger-action platform directly, users need to set up rules in our IoT platform and then the \mark{rules} can be correctly committed to the ledger and exported to third-party trigger-action platforms such as IFTTT after the transaction is confirmed.}

The ledger based IoT platform uses the pBFT~\cite{pbft} algorithm to perform the consensus process of \mark{recording updates}, which \mark{ensures} that the platform is secure unless \mark{more than} one-third of nodes committing the records are compromised. Note that, when the user account leaks, attackers can only exploit the devices under the single account but cannot perform vertical privilege escalation. In particular, all rules and settings are committed to the ledger, which makes the rules and devices configuration tamper-proof.

\subsection{Triggering Event verification}
{\ourwork prevents fake triggering events by verifying the execution logs of IoT devices and the event states recorded in the ledger. We utilize a ledger wallet called an execution agent to ensure that events recorded on the ledger are correct, which is used to protect device APIs and prevent privacy violation by verifying these records. }
%First, we explain how we achieve this goal by proposing an execution agent and why this design make it easier to protect the devices API and avoid privacy violation. Then we illustrate how the rules are triggered and how the execution log verification works.}

{Various execution agents in the IoT gateway execute different operations for the devices and commit the corresponding information to the ledger, which can be used to verify that all operations are correctly executed and APIs are invoked with correct OAuth tokens. According to our key observation that operations specified \mark{in} IoT rules are executed in separated procedures that \mark{run} in different processes or machines. Our execution agent can be integrated into the corresponding IoT gateway of the devices, and then the OAuth tokens of the device HTTP \mark{APIs} can be stored in the agent. Thus, the private data delivered by HTTP APIs will not leak \mark{into} the IoT platforms and the trigger-action platforms, and these platforms only obtain the execution results, instead of the user information, e.g., user locations in the plain text. Similar to the situational access control~\cite{situational}, for a specific device, we develop prefab scripts (see Appendix~\ref{adx:rule_setting}), that are similar to the code of SmartThings device handlers, which encapsulate the \mark{original} HTTP APIs to implement operations for the device. Note that, if a rule includes multiple operations, we need to split the rule such that each script is associated with one operation. Before an operation is executed, an execution agent associated with the device needs to pull the corresponding scripts from the ledger to perform the executions.}
%are proposed for each vendor's IoT gateway to execute the operations of their devices. This is because, in traditional IoT platforms, all the operations are executed in the platform and the OAuth tokens for the devices' gateway APIs need to store directly in the platform and the privacy data from the devices also can be accessed by the platform which may suffer attacks. Based on the observation that the operations of IoT rules are implemented by separate procedures that can be distributed into different processes or machines. We provide an execution agent module for the device vendors to integrate with their IoT gateway in their cloud service. In this way, the device's HTTP API OAuth tokens only need to be saved in the execution agent and the privacy data from the raw HTTP API do not leak to the IoT platforms and trigger-action platforms, e.g. to determine if the user is at home, we just share the result of whether the user is at home instead of sending user's location to the IoT framework. Similar to the situational access control proposed by~\cite{situational}, for a specific device, we provide prefab scripts (just as SmartThings' device handler) that encapsulate the origin HTTP APIs to implement the devices' operations and before a  operation is executed, the device's execution agent needs to pull the corresponding scripts from ledger. }

All trigger operations specified in rules are registered after the rules are set up. \mark{They} keep running in the execution agents that \mark{poll} the trigger state by executing the operation and \mark{check} the trigger conditions in a fixed interval (shown in Algorithm~\ref{alg:condition}). When the trigger condition in a rule is met, the execution agent \mark{generates} an event transaction that contains the rule information, the execution log's query key, and the corresponding checksum to the ledger. Upon receiving the information, the smart \mark{contract} verifies the transaction and the corresponding execution log. If the verification succeeds, the event transaction is committed to the ledger and the task agent will generate a trigger request\footnote{Note that most IoT platforms, such as IFTTT, support push mode and does not poll the trigger state. After the rule is triggered, the IoT platform pushs the triggering event to IFTTT.} to the trigger-action platform to trigger this rule (shown in Algorithm~\ref{alg:event}). \textsf{LedgerVerify} in Algorithm~\ref{alg:event} uses the event verification smart contract to verify the correctness of the transaction and stores it in the ledger if the transaction is authentic. 

\SetKwFor{For}{foreach}{\textbf{do}}{\textbf{endfor}}
\SetKwInOut{Input}{Input}

\begin{algorithm}[t]
\begin{footnotesize}
\caption{Check Trigger Condition}
\label{alg:condition}
\KwResult{Submit event transactions}
  // check the trigger conditions in execution agents\\
  \For{$R_i \gets Rules$}{
        $\textsf{trigger} \gets R_i $\\
        $\textsf{(eid, log\_key)} \gets \textsf{GenLogKey}(R_i) $\\
        $result \gets \textsf{ExecOperation(eid, trigger, log\_key)}$\\
        \uIf{$ \textsf{CheckTriggerCondition}(result, R_i) $}{
            // submit the event transaction \\
            $\textsf{log\_sum} \gets \textsf{CheckSum(}result\textsf{)}$\\
            $\textsf{event\_log} \gets \textsf{(eid, log\_key, log\_sum)}$\\
            $\textsf{event\_info} \gets \textsf{GenEventId}(R_i)$\\
            % $\textsf{VerifyLog(log\_key, log\_checksum)} $\\
            $\textsf{SubmitTransaction(event\_info, event\_log)}$\\
        }
  }
\end{footnotesize}
\end{algorithm}
%\vspace{-0.1in}
\begin{algorithm}[t]
\begin{footnotesize}
\caption{Triggering Event Verification}
\label{alg:event}
\Input{An event transaction, $T_i$}
\KwResult{Gen trigger requests}
  // verify the event transaction and send trigger requests\\
  $\textsf{(event\_info, event\_log)} \gets T_i$\\
  \uIf{$\textsf{LedgerVerify(event\_info, event\_log)}$}{
    %  $req$\\
     $\textsf{CommitTransaction(}T_i\textsf{)}$ \\
     $\textsf{Cid} \gets \textsf{GenRandomness(event\_info)}$\\
     // notify the task agent to send trigger requests\\
     $\textsf{TaskAgentSendRequest(event\_info, Cid)}$\\
  }
\end{footnotesize}
\end{algorithm}

\begin{figure}[t]
    \centering
    \includegraphics[width=0.45\textwidth]{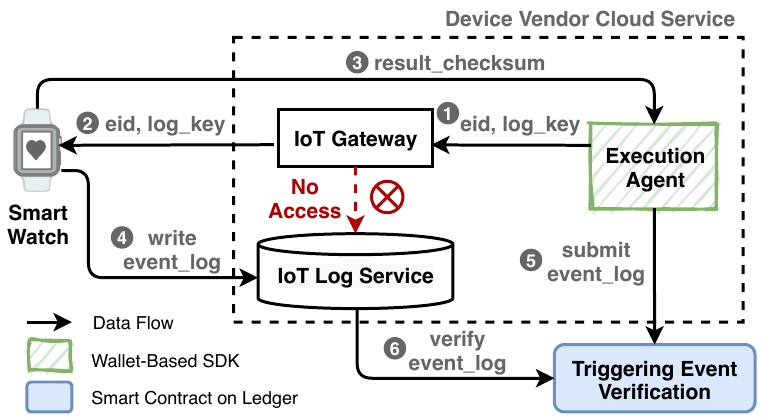}
    \vspace{-0.03in}
    \caption{The procedure of triggering event log verification}
    \label{fig:verify}
\end{figure}

{In order to defend against the \mark{event spoofing attacks} on the devices, we utilize an execution log based event verification mechanism. Before a record transaction is accepted by the ledger, the ledger will verify the authenticity of the record by querying the execution log associated with the event.  We leverage the existing logs generated by the device's cloud service and include a random key for each entry of the execution log, which ensures that only real devices can write the log. For a specific operation, the execution agent generates an execution id and a random key as the log query key pair (\textsf{eid, log\_key}). As is shown in \mark{Fig.}~\ref{fig:verify}, this log query key pair is also sent to the device through the gateway (steps \ding{202}, \ding{203}). If the operation is successfully executed by the device, a new \textsf{event\_log} that contains the log query key pair and the checksum of execution result is included \mark{in} the log (step \ding{205}). Also, the execution agent can obtain the same checksum (step \ding{204}) with the execution results and submit  the \textsf{event\_log} as a transaction (step \ding{206} (see Algorithm~\ref{alg:condition}). Finally, the triggering event verification smart \mark{contract} in the ledger checks if the \textsf{event\_log} associated with the transaction is the same as the record in the IoT log (step \ding{207}). As all valid device operations are initiated by the execution agent with the unique log query key and the IoT gateway cannot directly access the log, \ourwork can easily identify malicious executions triggered by the gateway or devices by checking the verifiable \textsf{event\_log}. }

{Therefore, the authenticity of triggering events can be ensured. Particularly, it is difficult for attackers to generate a fake event, which \mark{requires} compromising both the device and the gateway, or both the log server and the gateway. Note that, we do not consider collusion attacks among several parts, 
%are malicious, 
which means the attacker can bypass the whole \mark{verification system}.
%and can do anything they want. 
However, if APIs are manipulated or devices are compromised, the valid execution log associated with the operation execution will not be generated since the attacker does not have the key to generate the records, which is generated by the execution agents in the gateway.}
%which is generated by the execution agent for each operation.}

%\red{In this way, the authentic of the triggering event can be ensured and it is hard for the attackers to send a fake event which need to  compromise both the device and the gateway or both the log service and the gateway. Note that we do not consider these types of collusion attacks that several parts are malicious which means the attackers can bypass the whole systems and achieve their goals in many ways. If only the API services or the physical devices are compromised, the valid execution log can not be generated as lack the specific log accessing key which is generated by the execution agent for each operation.}

% The execution agent only need to protect its wallet private key and the OAuth token for the specific cloud service. We suppose the case when the OAuth token leaks, the attackers can use this token to send request to the cloud service and misuse the devices. With the overprivilege OAuth token attackers can still exploit the devices but they can not trigger the following actions of any rules, as other actions can only be triggered by the record transactions which is submitted by the execution agent. Without trust execution environment, the token misuse can not be avoided as there are so many different IoT cloud services and they all use OAuth-like protocol to control their devices. In our design, we do not require TEE and use the log service to avoid the fake event. 

\subsection{Verifiable Action Execution in \ourwork} 
%\red{Now we present how we address the challenges of trigger requests verification as the trigger-action platform is untrusted and may trigger overprivileged transactions. }
% The execution of the trigger-action rules involves multi-step interactions with the ledgers.  As is shown in Figure~\ref{fig:workflow}, the triggering event and action event are both need to submit to the ledger as transactions (steps \ding{204}, \ding{208}) via the wallet-based task agent and execution agent. The task agent is used to read the triggering event's transaction record  (step \ding{205}) and send the trigger request to the trigger-action platforms (step \ding{206}) and receive the action request from the trigger-action platforms  (step \ding{207}) and submit the action event as transactions to the ledger (step \ding{208}). The execution agent is used to read the action's transaction record form the ledger (step \ding{209}) and execute the action task (step \ding{210}) and submit the action's execution record as a transaction to the ledger which can be used as logs for rule execution history data analysis.  The challenges can be summarized as how to prevent invalid transactions triggered by the devices or devices gateway services or the trigger-action platforms. 
% To solve the problem, we build stateful event records on the ledger. The ledger interface is defined as,
% $$ Task(TaskInfo, Cid) \longrightarrow Task_{Cid} $$
% $$ Event(TaskInfo, ResultInfo, Cid) \longrightarrow Record_{Cid} $$
Now we describe how \ourwork realizes action requests verification, which aims to detect fake triggers and overprivileged transactions in the untrusted trigger-action platform. {In the current trigger-action platforms, all rules and devices of a user share the same OAuth token, which makes a malicious trigger-action platform arbitrarily \mark{generates action requests}. To address this issue, \ourwork utilizes fine-grained rule control on the trigger generation. %and request validate mechanism is needed. 
A triggering event is sent to the trigger-action platform only when the rule  trigger condition is satisfied. To distinguish different requests, we deliver the event identity \textsf{eid} and a random coin \textsf{Cid} that is generated by a pseudorandom function according to the event information. Note that, here we cannot use the existing token based solution~\cite{dtap18} that leverages \mark{per-rule} token for each triggering event, which \mark{cannot} ensure the authenticity of triggers generated from the trigger-action platforms.
%still incur OAuth leaks. 
%instead, the unique \textsf{Cid} can be used as the credential for each rule triggering. 
To trigger a rule, the trigger-action platforms need to pass the correct \textsf{eid} and \textsf{Cid} to the \mark{task agent}. Thus, attackers cannot forge the corresponding \textsf{eid} and \textsf{Cid} but only rewind the existing requests. However, after the actions are triggered, the execution records associated with the actions are also committed to the ledger by the execution agent. The rewinding requests will be rejected by the task agent and the ledger. Thus the malicious transactions are prevented and the attackers cannot exploit rules generated by the trigger-action platforms.}
%\red{In current trigger-action platforms, all the rules and devices of a single user share the same OAuth token and this makes a malicious trigger-action platform can arbitrarily send action trigger requests. To solve the problem, fine-grained rule control and request validate mechanism is needed. In our system, the triggering event contains is sending to the trigger-action platform when the rule's trigger condition is satisfied by the task agent. To distinguish different requests, we pass the event identity \textsf{eid} and a randomness coin \textsf{Cid} which is generated by a pseudorandom function based on the event detail. We do not propose token-based solution such as 
%DTAP~\cite{dtap18} which uses pre-rule token for each triggering event, instead, the unique \textsf{Cid} can be used as the credential for each rule triggering. To trigger a rule, the trigger-action platforms should pass the correct \textsf{eid} and \textsf{Cid} to the user agent. The attackers can't forge the corresponding \textsf{eid} and \textsf{Cid} and can only rewind the existing requests. However, after the actions are triggered, the actions' execution records are also committed to the ledger by the execution agent. The rewinding requests will be rejected by the task agent and the ledger. Thus the malicious transactions are prevented and the attackers cannot exploit rules from the trigger-action platforms.}

\begin{algorithm}[t]
\begin{footnotesize}
\caption{Action Verification}
\label{alg:trigger}
\Input{A action transaction from the task agent, $T_i$}
\KwResult{Submit action transactions}
  $\textsf{(event\_info, Cid)} \gets T_i$\\
  \uIf{$\textsf{VerifyRandom(event\_info, Cid)}$}{
        $\textsf{event\_record} \gets \textsf{LedgerQueryEvent(event\_info)}$\\
        \uIf{$\textsf{LedgerVerify(event\_record)}$}{
            $\textsf{rule} \gets \textsf{LedgerQueryRule(event\_info)}$\\
            // submit action operations as transactions \\
            $\textsf{SubmitTransactions(rule.actions)}$\\
            // the execution agent execute the actions and record the execution log \\
            $\textsf{NotifyExecutionAgent(rule.actions)}$\\
        }
  }
\end{footnotesize}
\end{algorithm}

{The pseudo-code of \mark{action verification} is shown in Algorithm~\ref{alg:trigger}. After the ledger receives a request of an action transaction, the action verification smart \mark{contract} (i.e., \textsf{LedgerVerify}, see Appendix~\ref{adx:smart_contact}) decides if the transaction should be committed and executed. Unlike the existing smart home systems where the rule execution condition is only checked by the trigger-action platforms, in \ourwork, the execution of an action operation is confirmed according to the event record in the ledger. Thus, attackers cannot arbitrarily trigger rules unless the trigger condition is met and the event transaction has been committed to the ledger. Moreover, with the tamper-proof event records in the ledger, we can ensure that such rules can only run once \mark{according to} the matched triggering event in the ledger. }

\subsection{Discussion}
{ %and may have much flexible complex logic. 
\ourwork ensures the integrity of rule execution in the trigger-action based smart home systems that use conditions as triggers to execute operations specified in rules. \ourwork %And thus, our system 
can validate all these conditions according to the state records in the ledger.} 
%\red \ourwork 
Thus, it does not aim to ensure the execution integrity directly triggered by SmartApps that are implemented in Groovy scripts. %, if it is integrated into the SmartThings.
{Moreover, the wallet based agents in \ourwork use \mark{different private keys} that should be protected by the existing key protection solutions~\cite{key_protect1, key_protect3} or the trusted execution environments. } %to protect keys and sign the transactions. }

%% file: sections/4-security-analysis.tex
\section{Security Analysis}
In this section, we analyze the security of \ourwork against the  attacks. 
%and clarify our security assumptions and thread model. We discuss how the API level attacks and the platform or devices compromised attacks affect our system and how to prevent malicious behaviors under these attacks. 

%\subsection{API Level Attacks}
\noindent \textbf{API Level Attacks.}
% As all the requests from the devices and the trigger-action platforms need to be submitted as ledger transactions, we . 
\ourwork prevents API level attacks by ensuring the authenticity of users, rules, events, and triggers, and thus guarantees the execution integrity of smart home systems. %As we enable a ledger service in \ourwork, 
All functionalities are performed via transactions that are executed and verified by smart \mark{contracts}, and all execution records are stored in the ledgers. Thus, all fake events and fake actions triggered by APIs can \mark{be} detected  %since these records cannot be verified 
by \ourwork.

First, user accounts and rule configuration cannot be faked and manipulated. \ourwork utilizes a wallet private key to sign user accounts and rule configuration in the configuration synchronization phase, and \mark{records} these information into the ledger via a consensus process. % it totally  
Any attacker cannot abuse API to inject fake users or rule configurations due to the lack of the key. Thus, all \mark{users} and rule information cannot be faked and manipulated by API level attacks. %if the private key does not leak. 
%Here we only discuss the attacks to the task agent and execution agents to trigger malicious transactions. We need to consider the task agent API security as well as the execution agent API security.
%all the functionalities are implemented via transactions that are executed and verified by the smart contacts, and finally, the records are stored in the ledgers. For the user account and rule configuration, it totally relies on the wallet-agent private key security and the ledger consensus which cannot be violated by API-level attacks if the private key does not leak. Here we only discuss the attacks to the task agent and execution agents to trigger malicious transactions. We need to consider the task agent API security as well as the execution agent API security.

Second, \ourwork  ensures that any events in smart home devices actually occur. Particularly, the execution of the operations can only be triggered by the trigger transactions on the ledger via the execution agents of \ourwork and each operation associated with event changes can only be triggered once through the \mark{execution agent}. Thus, all execution logs associated with events are recorded in \ourwork. The attacks constructed by any malicious users, e.g., via a compromised user agent, cannot abuse any rule or trigger malicious execution of an operation specified in a rule because they cannot generate fake triggers that can be verified by \ourwork. %generated from the %form the user agent's side can not exploit any rules or a single operation of the rule. To exploit a rule, the attackers need to send a fake trigger event to the execution agent and compromise the log-record service.
%We ensure that the execution of the rule can only be triggered by the trigger transactions on the ledger from the execution agents and each operation can only be triggered once from the task agent. The attacks form the user agent's side can not exploit any rules or a single operation of the rule. To exploit a rule, the attackers need to send a fake trigger event to the execution agent and compromise the log-record service.
%In our design, we can prevent all the API level attacks if the wallet-agents' private keys do not leaks.
% For the devices API security, we assume that 

Third, triggers in \ourwork  based smart home systems cannot be faked so that requests generated from the trigger-action platforms 
are authentic. 
%The task agent may receive malicious requests from the trigger-action platforms via rewind attacks or token leaks. 
In order to execute a rule in \ourwork, different parameters, e.g., event info and a temporary key \textsf{Cid}), are submitted to the trigger-action platform, and then the platform generates trigger requests that carry these parameters to trigger specific actions for IoT devices. As rule configurations and new rule execution tasks under execution are triggered by the execution agent of \ourwork in the devices, an attack against the trigger-action platforms can only exploit a single operation by generating and submitting fake parameters. Even if the attacker can inject the correct parameters to trigger the execution of an operation, the ledger in \ourwork can ensure that the operations of a task with correct parameters can only be executed once and the \mark{action} operations are really executed based on the trigger records in the \mark{ledger}. Note that, the trigger-action platform can only trigger an operation with valid parameters once \mark{when} the trigger event is authentic and the operation is not overprivileged.
%The task agent may receive malicious requests from the trigger-action platforms via rewind attacks or token leaks. To run a rule, several parameters (task info and a temp key \textsf{Cid}) are passed to the trigger-action platform and the platform needs to send trigger requests which carry these parameters to trigger specific actions. As the rule setting and new rule execution task submitting are motivated by the execution agent's side, the attacks from the trigger-action platforms can only try to exploit single operations by sending fake parameters. Even if they can feed the correct parameters to trigger the execution of an operation, the ledger can ensure that the operations of a task can only be executed once and whether the actions operations will really be executed based on the ledger's trigger records. It means that the trigger-action platform can only trigger an operation once with the valid parameters only when the trigger event is real and the operation is not overprivilege.

%\subsection{Platform Compromise Attacks}
\noindent \textbf{Platform Compromise Attacks.} 
% attacker that can obtain root access to devices (e.g., Mirai attack [10]),
\ourwork can prevent the platform compromise attacks that may be constructed in each component. We will show how \mark{the} execution integrity is ensured even under different platform compromise attacks.  %try to explore how to ensure the rule execution integrity with the least security assumptions and analyze the attack effects to each parts.

If the trigger-action platform is compromised, attackers still cannot abuse the rules of various users to generate malicious requests because \ourwork will filter out such requests that are not recorded in the ledger of \ourwork.  
%and lead to a larger scale API exploiting. This kind of attack and malicious requests can be prevented by our ledger transactions.
%If the trigger-action platform is compromised, the effect is that attackers can exploit all the rules of various users and lead to a larger scale API exploiting. This kind of attack and malicious requests can be prevented by our ledger transactions.
If \mark{the ledger} based IoT platform is compromised, \ourwork can still ensure the rule execution integrity.  
%need to consider the security of the ledger and the wallet security. 
Since rules and the corresponding operation records are triggered by smart contracts and generated via \mark{ledger} transactions, the ledger nodes verify and execute the transactions by using the pBFT consensus algorithm~\cite{pbft} that can tolerate at most one-third of nodes being compromised. In the worst case, one-third of nodes are malicious, \mark{and they could prevent the committing} of transactions or submit malicious \mark{transactions}. The ledger in \ourwork can still reach consensus and commit the transactions successfully.
Note that, if devices \mark{or} the corresponding gateway are compromised, we cannot prevent the abuse of compromised devices. However, \ourwork can still prevent the attack of event spoofing since the attacker \mark{cannot} inject fake logs of operation execution. The reason is that any fake events cannot produce fake logs in the log service of \ourwork as they are not invoked by the execution agent and cannot obtain the random log key to record the events.   
%are not invoked by the execution agent and can not get the randomness log key}.  
%If there are no colluding attacks to the log services and execution agents, the attackers cannot inject fake execution logs and the event spoofing attacks can be prevented.
%If devices or(/and) the corresponding gateway are compromised, we cannot prevent the abuse of compromised devices. If there are no colluding attacks to the log services and execution agents, the attackers cannot inject fake execution logs and the event spoofing attacks can be prevented.

%% file: sections/5-implementation.tex
\section{Performance Evaluation}
\subsection{Experiment Setup}
We prototype \ourwork by using an open-source blockchain system \cite{code}. %\footnote{https://github.com/FISCO-BCOS/FISCO-BCOS}.
The task agent, execution agent and user agent are implemented as web services with the Python Django framework invoking the wallet SDK to commit transactions and rules onto the ledger.
%As discussed in Section~\ref{sec: design}, we split the rule into multiple trigger scripts if the rule includes more operations. For example, if there is a rule \textit{"if the user's heart rate is serious abnormal, then open the smart lock of the door to let the ambulancemen in"}, the operations can be split into two Python scripts, i.e., \textit{"query the heart rate and check if it breaks the threshold"} and \textit{"open the smart lock"}. 

We evaluate the performance of each \ourwork module and the end-to-end system performance. In particular, we measure the latency and throughput of \ourwork \mark{modules} and \mark{the} integration \mark{of Ruledger} with SmartThings and IFTTT. The prototype system is deployed in seven elastic cloud servers and each \mark{is} configured with 4 core Intel Xeon CPU (3.10 GHz), 8G memory, and Ubuntu 20.04 OS. There are 4 blockchain nodes deployed on 4 servers. Note that in order to measure the throughput, we need to send multiple requests concurrently, which is constrained by the rate-limit mechanism in the SmartThings Device Handler and the IFTTT. Thus, we implement a skeleton \mark{device}  simulator and a trigger-action service similar to IFTTT Maker to measure the throughput of the system with and without \ourwork.

% Since the smart home devices on the market have limitations on remote API calls, for better evaluation of the key mechanism we simulate the devices by SmartThings' simulators~\cite{smartthings-web} and ourselves.

\subsection{The Performance of \ourwork Modules}
The main overhead of \ourwork is incurred by verifying rule triggers and \mark{rule} execution,  which are performed via smart contracts and \mark{the corresponding} requests are converted to transactions. 
%, which may impact the overall performance of \ourwork. the latency and throughput of the modules interacting with the ledger. 
In this experiment, we measure the performance of the two modules and their impacts on the platforms. 
%, which is considered the performance bottleneck \mark{of \ourwork}. 
\red{To accurately measure the performance, we use the task agent and execution agent to perform mock tests for transaction commit\mark{, and these} agents do not interact with devices and \mark{trigger-action} platforms but directly commit the mock transactions to the ledger. Note that, the delay of the task agent and the execution agent includes the delay incurred by other modules, such as request parsing and operation executions, which \mark{could} have various delays corresponding to the detailed rules.}

\noindent \red{\textbf{The Latency Incurred by Modules.}}
\red{To measure the latency of the transaction commit \mark{which is incurred by the} trigger event verification smart contract module and the action verification smart contract module, we use an execution agent and a task agent to submit test transactions to the ledger directly. These transactions do not need to be triggered by requests from device \mark{gateways} or trigger-action platforms so that we can get the real transaction commit time without the network latency. The average latency of the trigger event verification module is 32.45ms which includes all the procedures from \mark{a} transaction \mark{being} submitted by the execution agent to the verification of the smart contract and \mark{the transaction} finally \mark{being committed} to the ledger. And this time for the action verification module is 32.83ms. As the execution of a rule needs to commit the two transactions, the total latency of the smart \mark{contract modules} is less than 70 ms which is acceptable as it is only 4.36\% of the whole rule execution \mark{latency}.}

\noindent \red{\textbf{Throughput of Smart Contract Modules.} \mark{Since} a transaction \mark{is} committed only when all the ledger nodes reach consensus via pBFT~\cite{pbft}, the number of transactions that can be committed by the ledger in one second is limited. We submit transactions concurrently to test the ledger's transaction throughput threshold \mark{and} check if it is the \mark{performance} bottleneck of rule executions. We use the task agent and execution agent to submit different numbers of concurrent transactions and check how many transactions can be committed in a limited time.}

\begin{figure}[t]
\begin{subfigure}{0.234\textwidth}
\includegraphics[width=\linewidth]{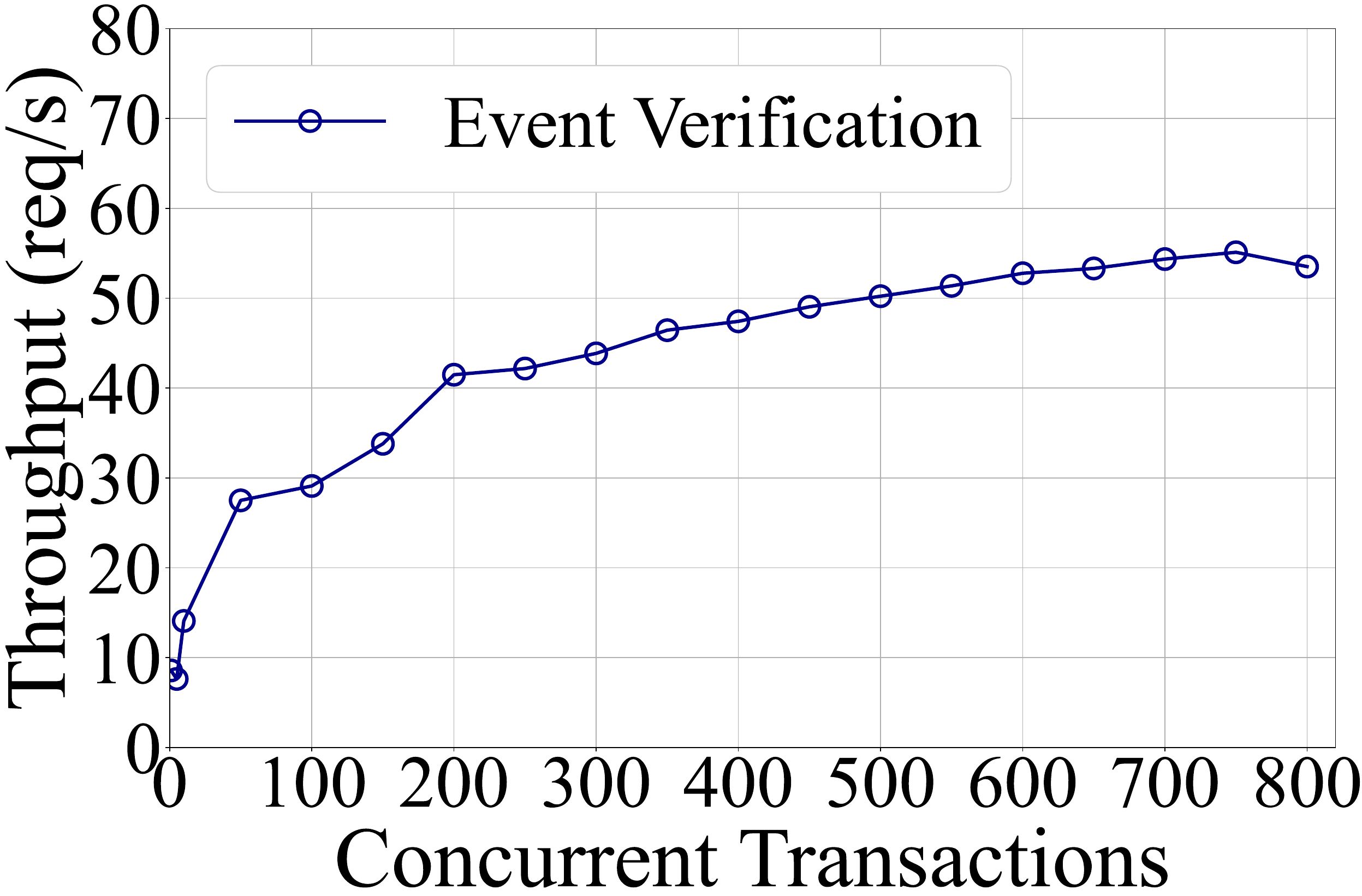}
%\vspace{-0.5in}
\caption{Triggering Event Verification} \label{fig:t1}
\end{subfigure}
\hspace*{\fill} % separation between the subfigures
\begin{subfigure}{0.234\textwidth}
\includegraphics[width=\linewidth]{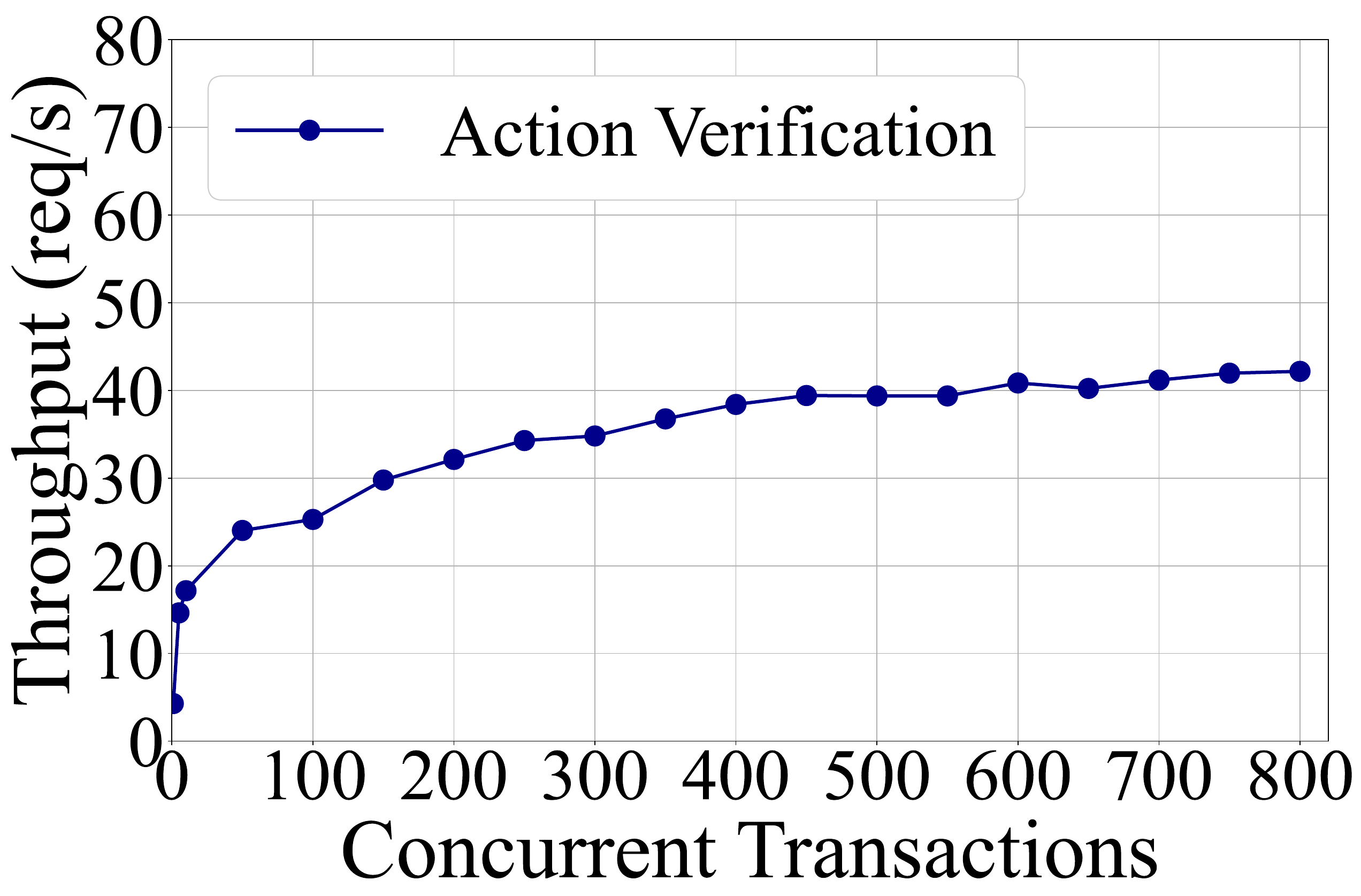}
%\vspace{-0.5in}
\caption{Action Verification} \label{fig:t2}
\end{subfigure}
%\vspace{-0.5in}
\caption{The throughput of Ruledger verification modules} \label{fig:smarthome}
\end{figure} 

% \begin{figure}[!htb]
% \begin{tabular}{cc}
% \begin{minipage}[t]{0.5\linewidth}
%     \includegraphics[width = 1\linewidth]{figrues/thea.eps}
%     \caption{The throughput of trigger event verification module with 4 nodes.}
%     \label{fig:t1}
% \end{minipage}
% \begin{minipage}[t]{0.5\linewidth}
%     \includegraphics[width = 1\linewidth]{figrues/thta.eps}
%     \caption{The throughput of action verification module with 4 nodes.}
%     \label{fig:t2}
% \end{minipage}
% \end{tabular}
% \end{figure}

The throughput is plotted in \mark{Fig.}~\ref{fig:smarthome}. 
%~\ref{fig:t1} and \ref{fig:t2}.
The concurrent transactions are submitted to the modules with valid test data \mark{for verification}. By recording the start time and finish time of the transactions, we can measure the number of transactions processed per second. The throughput of the triggering event verification module is 43 TPS (transactions per second) and \mark{that of} the action verification module is 55 TPS. Also, we evaluate how ledger nodes affect the throughput. The results show that the throughput is not significantly impacted by \ourwork. % with 10 nodes or less. 
The results demonstrate the feasibility of \ourwork. %modules for real deployment.
%are viable for the application of a real-world smart home system, 
%\mark{let alone the optimizations in the industrial deployments}.

\subsection{Performance of The Entire System}
\red{We measure \ourwork's overhead of rule execution \mark{in real-world deployments} by integrating it with SmartThings and IFTTT \mark{and then testing} the rule execution latency \mark{as well as} the throughput of the whole system. We use the IFTTT Maker channel to trigger the \textit{"heart rate alert"} rule and set \mark{simulated} devices for a smart watch and a smart lock in the SmartThings WebIDE~\cite{smartthings-web}.  }

% To make it easier to trigger rule, we use two simulate devices for the smart lock and the smartwatch in SmartThings WebIDE~\cite{smartthings-web} so that we do not need to connect the SmartThings's Device handler to the real devices of smartwatch and smartlock. 
% However, for the throughput test which needs to send concurrent requests,  both the IFTTT maker and the SmartThings Device Handler have request rate limited. We implement our own simulator devices for automatically rule concurrent executions and a custom trigger-action service which works the same as IFTTT Maker to process unlimited requests.

\red{\noindent \textbf{End-to-End Latency.}}
\red{To measure the latency of a round-trip \mark{from} rule triggering \mark{to} action execution, we deploy \ourwork as a middleware between the SmartThings platform and IFTTT \mark{and then execute rules under the protection of this middleware.} And the baseline is \mark{implemented using} the default usages of IFTTT that IFTTT directly controls the \mark{device handlers of} SmartThings and do not use \ourwork as a middleware. We record the whole time of \mark{triggering} and execution of the \textit{"heart rate alert"} rule \mark{that is required by} our ledger \mark{based} version and the \mark{baseline}.}

\begin{table}[t]
\centering
\caption{Comparison of average execution latency.}\label{tbl: overall}
%\vspace{-0.08in}
\begin{tabular}{cccc}
\hline
              & SmartThings & \ourwork & Delay \\ \hline
End-to-end execution latency \\ in average of 30 trials & 1.403 s    & 1.604 s   & 12.53\%  \\ \hline
\end{tabular}
\label{tab:latency}
\end{table}

The result is shown in Table~\ref{tab:latency}. The execution latency \mark{of the baseline} includes \mark{time for} two requests, the event request from \mark{a smart watch simulator of SmartThings} to the IFTTT Maker channel, and the \mark{action request} from the IFTTT to \mark{a} smart lock device. \mark{By using logs in} the SmartThings WebIDE, we can record the event request's sending time as the start time and the \mark{action} request's \mark{receiving} time as the end time to get the round trip time. However, it is impossible for \mark{users} to filter out the network latency and the IFTTT's processing time \mark{since} we can not get any log \mark{with an accurate time of the processing} from the IFTTT. In the \mark{baseline system}, the execution of the whole rule finished in 1.403 s and in the \mark{system based on} \ourwork, this time is 1.604s which includes \mark{time for interactions with the ledger} and the \mark{time of sending} requests. The overall latency \mark{of} rule execution \mark{shows a 12.5\% increase in the \ourwork version which is about 200 ms, comparing with the baseline}.

\red{The IFTTT's rule trigger~\cite{ifttt_study} is not generated in real time as the default polling mode has a delay of seconds or minutes and the pushing mode incurs a delay of less than 1s. The latency of \ourwork does not affect the rule execution as the IoT rules do not need to be triggered immediately, which can tolerate a delay of several hundreds of milliseconds.}

\red{\noindent \textbf{Throughput.} As the SmartThings \mark{device simulators} and the IFTTT have rate limit and cannot \mark{be triggered} in a high frequency, we implement two \mark{simulated} devices for smart watch and smart lock which can return the heart rate data or execute the unlock operation via HTTP requests\mark{, and we also implement} a custom trigger-action service which works the same as IFTTT Maker to process unlimited requests. \mark{The rule could} be triggered \mark{by sending requests with an abnormal heart rate,} and we adjust the request number and frequency to test the throughput. For the \ourwork version, we use the execution agent and the task agent to convert requests to transactions and verify these \mark{transactions} via ledger and smart contracts. In the baseline system, we just let the execution agent send origin requests to the task agent and do not interact with the ledger, as well as \mark{letting} the task agent directly send the \mark{action requests} to the execution agent to execute actions without the ledger's verification. In this way, we ensure \mark{that} the network and machine status are the same in the two deploy versions.}

\begin{table}[t]
\centering
\caption{The throughput of \ourwork and SmartThings with 2000 concurrent requests} \label{tbl: through}
%\vspace{-0.08in}
\begin{tabular}{cccc}
\hline
    & Baseline & \ourwork & Decrease \\ \hline
    Throughput (req/s) & 43.37  & 40.57  & 6.5\% \\ \hline
\end{tabular}
\label{tab:throughput}
\end{table}

To measure the throughput of rule processing, 
%how many rules can be processed in limited time, 
we construct 2,000 concurrent requests in the  experiment. As shown in Table~\ref{tab:throughput},  the system without ledger processes 43.37 requests of rule execution per second, and \ourwork handles 40.57 requests per second. The throughput is reduced by 6.5\%, which is negligible.

%% file: sections/7-discussion.tex
\section{Related work}
\noindent \textbf{Trigger-action Platform Security.} DTAP \cite{dtap18} aims to provide action integrity for rule execution, which is most related to our work. It can prevent arbitrary rule executions \mark{from exploiting} leaked OAuth tokens. In particular, it leverages coarse-grained XTokens in devices to protect rule-specific tokens, and \mark{verifies} rule executions for action services. %according to the trigger event record in the API call and trigger service mapping in the action token. 
Unfortunately, it \mark{is unable to} prevent attacks that \mark{use vulnerable APIs to generate fake events}. \mark{Besides, DTAP can only constrain the ability of an attacker to steal a rule-specific token rather than preventing the attacker from exploiting rule execution}. %by using leaked rule-specific tokens. 
\ourwork well addresses this issue. \ourwork utilizes the \mark{tamper-proof} feature of the ledger to verify the authenticity of the information and \mark{the} integrity of rule execution. Moreover, %most of existing trigger-action platforms do not provide event verification for smart home systems. However, 
%prior studies~\cite{peeve} have shown that attackers could deceive the platforms with fake events, which is called 
The event spoofing attack\cite{fernandes2016security} and active attacks\cite{wang2019charting,zhang2019autotap,celik2018soteria,celik2019iotguard,chi2018cross}  against trigger-action platforms~ is not well addressed in the literature. %Moreover, the chain of trigger-action rules are vulnerable to various active attacks \cite{wang2019charting} \cite{zhang2019autotap} \cite{celik2018soteria} \cite{celik2019iotguard} \cite{chi2018cross}. 
Our \ourwork well addresses these issues by automatically recording verifiable operation execution records in ledger with smart contracts. 

%  and IoTivity \cite{IoTivity}
\noindent \textbf{Access Control for IoT System.}
IoT applications in IoT framework, e.g., SmartThings \cite{smartthings-web}, use OAuth tokens to execute rules in target IoT cloud \mark{services}, which suffers from the overprivileged issue due to the coarse-grained access control and lack of functionality isolation. Moreover, the exposure of original device APIs may incur privacy leakage. A number of studies focus on developing fine-grained access control mechanisms for IoT. For instance, FlowFence \cite{fernandes2016flowfence} utilizes a secure flow control mechanism to control sensitive data \mark{used by} apps by executing device functions in sandboxes. FACT \cite{lee2017fact} uses access control list (ACL) registered by user to examine the privileges of applications.  %attempting to execute rules. %It uses the virtualization technique to isolate different resources to protect IoT devices from DoS and side-channel attacks. 
Situational access control \cite{situational} provides function wrapping for device APIs and uses unified IoT events, instead of revealing privacy data to IoT framework platforms or trigger-action platforms. These approaches can prevent \mark{overprivileged} rule executions. Although these mechanisms heavily rely on OAuth tokens, %are burdened with protecting device handlers and their executions for each user. Moreover, 
device APIs can be easily abused if OAuth tokens are leaked. In order to prevent violation of execution integrity, \ourwork utilizes the device scripts in the ledger to verify the authenticity of the information, instead of using a token to execute rules.

\noindent \textbf{Blockchain for IoT Security.}
Blockchain has been leveraged to protect IoT~\cite{minoli2018blockchain} \cite{novo2018blockchain} \cite{derler2019fine} \cite{ouaddah2016fairaccess} \cite{maesa2019blockchain}. 
%Many researchers have studied security protection for IoT system using blockchain \cite{minoli2018blockchain}. 
For example, a blockchain based smart home framework \cite{minoli2018blockchain} % It uses a local private blockchain 
records device management and device communication history as time ordered transactions and enforce these transactions according to embedded policies. Moreover, %Some 
several blockchain based access control approaches~\cite{tang2019iot, novo2018blockchain, derler2019fine, ouaddah2016fairaccess,  zyskind2015decentralizing, maesa2019blockchain} have been proposed. Our \ourwork is orthogonal to these approaches. We can leverage these approaches to implement fine-grained access control in the ledger.

%% file: sections/9-appendix.tex
\section{Rule and Smart Contract Samples in \ourwork} 
\subsection{A Sample Smart Contract for Information Verification}
\label{adx:smart_contact}

\definecolor{codegreen}{rgb}{0,0.6,0}
\definecolor{codegray}{rgb}{0.5,0.5,0.5}
\definecolor{codepurple}{rgb}{0.58,0,0.82}
\definecolor{backcolour}{rgb}{0.95,0.95,0.92}

\lstdefinestyle{mystyle}{
    backgroundcolor=\color{backcolour},   
    commentstyle=\color{codegreen},
    keywordstyle=\color{magenta},
    numberstyle=\tiny\color{codegray},
    stringstyle=\color{codepurple},
    basicstyle=\ttfamily\footnotesize,
    breakatwhitespace=false,         
    breaklines=true,                 
    captionpos=b,                    
    keepspaces=true,                 
    % numbers=left,                    
    numbersep=5pt,                  
    showspaces=false,                
    showstringspaces=false,
    showtabs=false,                  
    tabsize=2
}

% pragma experimental ABIEncoderV2;
% import "./ORM.sol";
% import "./LibOwnable.sol";
% import "./LibUtils.sol";

\lstset{language=Java, style=mystyle}
\begin{lstlisting}
pragma solidity ^0.4.24;

contract Executor is LibUtils, ORM{
  function ledgerVerifyTrigger(int usr_rule_id, int 
  usr_id, int rule_id, string rule_name, int step_id) public returns(int) {
    // check rule execution access
    int memory usrItem = verifyUsrRule(usr_rule_id, usr_id, rule_id, rule_name);
    if (usrItem == RES_ERR) {
      return ERR_USER_VERIFY_FAILED;
    }
    // triggering event verification
    string[] memory conKeys = new string[](2);
    int[] memory conVals = new int[](2);
    conKeys[0] = "tRule_id";conKeys[1] = "tStep_id"; 
    conVals[0] = usrItem; conVals[1] = step_id - 1;
    string[] memory trEvts = new string[](3);
    trEvts[0] = "tTask_id"; trEvts[1] = "tStep_id"; trEvts[2] = "tResult";
    int[] memory rets = selectEntry(trEvtTblName, conKeys, conVals, trEvts);
    // check if the trigger record exist
    if (rets.length <= 0 || rets[1] != step_id - 1 || rets[2] != RES_OK) {
      return ERR_TRIGER_VERIFY_FAILED;
    }
    return RES_OK;
  }
}
\end{lstlisting}

\subsection{Defined Rules in \ourwork}\label{adx:rule_setting}

\lstset{language=Python, style=mystyle}
\begin{lstlisting}
def alert_on_heart_rate(device_id, token):
	data = RPC_CALL(HEAR_RATE_API, device_id, token)
	if data.success:
	    heart_rate = data.get("heart_rate")
	    return heart_rate <= MIN_RATE or
	        heart_rate >= MAX_RATE
	return False

def open_door_operation(device_id, token):
	res = RPC_CALL(SMARTLOCK_UNLOCK, device_id, token)
	return res.success

title = "alert on heart rate"
TRIGGER_OPERATIONS = [(alert_on_heart_rate, device_info1, OP_AND)]
CONDITION = IF_TRUE # lambda v: v is True
ACTION_OPERATIONS = [(open_door_operation, device_info2, OP_AND)]
RULE_DEFINE(title, TRIGGER_OPERATIONS, CONDITION, ACTION_OPERATIONS)
\end{lstlisting}